\documentclass[11pt]{article}
\usepackage{amsmath,amssymb}
\usepackage{graphicx}
\usepackage{braket}

\usepackage{enumerate}

\newcommand{\bs}{\boldsymbol}

\begin{document}
\title{Topological interaction of neutrino with photon \\
in a magnetic field -- Electroweak Hall effect-- }    
\author{
{Kenzo Ishikawa} $^{(1)}$ and  { Yutaka Tobita} $^{(2)}$}
\maketitle
\begin{center}
(1) Department of Physics, Faculty of Science, \\
Hokkaido University, Sapporo 060-0810, Japan \\
(2) Department of Information and Computer Science, Faculty of Engineering,
Hokkaido University of Science, Sapporo 006-8585, Japan \\\end{center}


\begin{abstract}
The effective interaction of a neutrino with a photon in magnetized plasma is obtained from a strong field expansion in 
 the electroweak standard model.  The interaction is expressed by a Chern-Simons form of the neutrino current and the electromagnetic vector potential of the coupling strength proportional to
 $\frac{n_e}{B} \times e G_F$. The derivation of the interaction Lagrangian 
 and its properties are presented.

\end{abstract}
\newpage
 \section{Introduction}
{ \bf Neutrino magnetic moment}

There are three types of neutrinos, and they  have spin $\frac{\hbar}{2} $, no electric charge, and  small masses. In the electroweak theory,  
these  belong to  $SU(2) \times U(1)$ doublets.  They  do not interact with a photon at at classical levels, and higher-order effects of the electroweak gauge theory  induce this  interaction. These effects are due to fluctuations of charged fields with masses, and the interaction in the vacuum occurs at a short distance.  
A neutrino turns to have a magnetic moment due to  higher order corrections due to   an induced  neutrino  energy  in a  vacuum with the external magnetic field $B$.
 This energy   is expressed as a series of positive power $B^n;n=0,1,\cdots$ on the strength of the magnetic filed $B$. The neutrino magnetic moment  is     extremely small  due partly to  tiny neutrino mass in the standard theory \cite{Raffelt, Fujikawa-Shrock}.

Calculations of higher-order corrections using an electron propagator in a vacuum preserve manifest Lorentz invariance and make the effective Lagrangian of an external vector potential manifestly invariant. A magnetic moment in a system of matter with electron state of momentum  smaller than the Fermi momentum, which is at most keV$/c$, is estimated similarly. One particle states above this energy remain empty and do not contribute. An effective coupling strength of the weak interaction in this region is $G_F$, an order of $\frac{1}{M_W^2}$. The cross-section for the typical energy $E$ is proportional to $G_F^2 E^2$, and minute for $E^2=2 M_{atom} E_{\nu}$ with a  small $E_{\nu}$.
The effect of matter on the neutrino magnetic moment is tiny when an electron propagator is  in the vacuum.   

The small couplings of the neutrino-photon interaction of these calculations are consistent with the Landau-Yang theorem, which asserts that the transition rate of the 
neutrino-photon reaction is suppressed irrespective of the details of the interaction \cite{landau,yang,gell-mann,particle-data}. We should note that this theorem concerns the rates, which satisfy naive conservation laws. As shown in \cite{ishikawa-tajima-tobita}, an additional term in the probability computed from the normalized states, the correction to  Fermi's golden rule, does not satisfy the theorem. It is worthwhile to study effective interactions that  may contribute to the rigorous transition probability \cite{ ishikawa-tobita-PTEP, ishikawa-tobita-ann}.  

Now, in a system of finite electron density in addition to the external magnetic field, the electron's classical orbit is closed  and very different from free motion, as expressed by the Landau levels in  quantum mechanics. Fluctuations of the electrons are described by a propagator of the electrons in the Landau levels, which  becomes substantially modified.

{\bf Quantum Hall effects}

A significant feature of finite electron density in external electric and magnetic fields appears in electric conductance.  Electric current flows in a perpendicular direction 
to electric and magnetic fields  due to the Lorentz force. Hall conductance has a signature of charge carrier density and is useful for detecting this information in metals.  Quantum effects become prominent when the Larmor radius $R$ is smaller than the system's size.  Intriguing quantum phenomena with semiconductors in a high magnetic field were 
observed in a system that satisfies this condition. A quantum  Hall effect with a Hall conductance quantized as $\frac{e^2}{4 \pi} N$ or  $\frac{e^2}{4 \pi} \frac{p}{r}$, where $N$, $p$, and $r$ are integer, was discovered.  At a small integer of $N$, intriguing Hall effects were observed in millimeter-long semiconductor chips in a Tesla-order magnetic field 
and at a temperature of 1 K \cite{klitzing}. 

The  quantization of the Hall conductance has been demonstrated in  various viewpoints. One  gave a rigorous proof of the exact quantization based on the general principles of quantum field theory. The Hall conductance was expressed by a topologically invariant quantity of electron propagator in a magnetic field, including interaction effects \cite{ishikawa-matsuyama}. This expression  is valid   in systems where a universal relation between a propagator and a vertex part known as Ward-Takahashi identity is satisfied. 
The derivation follows the fundamental principles of  quantum mechanics such as commutation relations, Schr\"{o}dinger  equation, and probability principle, and can be applied to a system where  Ward-Takahashi identity holds. In the electroweak gauge theory, $SU(2) \times U(1)$ symmetry is broken to  $U(1)$ by Higgs mechanism.  Electromagnetic  current is conserved,   and Ward-Takahashi identity is satisfied. Accordingly,  the quantum Hall  effect can be generalized to a system of neutrino and photon, which we call  an electroweak Hall effect. A neutrino has extremely small masses, and its sizable interaction with a photon may give new physical implications. 

Neutrino is neutral and does not couple with a photon in classical level. An interaction is induced from  higher order corrections. Electroweak Hall interactions  in magnetized plasma is different from the one  in a vacuum, which is known to be extremely small, and has a sizable magnitude.     
 
This paper shows that an anomalous term induced by  electrons in the magnetic field appears in a neutrino-photon interaction.  Using the electron propagators in the magnetic fields, the effective Lagrangian in the standard  electroweak gauge theory was computed from one loop corrections.  The included interactions reveal the magnetic field and have  Hall effects and an anomalous neutrino-photon interaction similar to the quantum Hall effect. An effective action has a Chern-Simons form of a coupling strength proportional to $\frac{\rho_e}{B}$, where $\rho_e$ is the electron density. This behavior is due to the quantum effects of electrons in closed orbits, which are specific to  the magnetic field and a finite electron density.

The obtained electroweak Hall interaction induces a long distance correlation similar to \cite{EPR}. This is independent of the short distance correlation and gives a correction 
for  Fermi's golden rule, leading to new physical phenomena \cite{ishikawa-tajima-tobita, ishikawa-tobita-PTEP, ishikawa-tobita-ann}, \cite{ishikawa-oda,ishikawa-nishiwaki-oda1,ishikawa-nishiwaki-oda2,Ishikawa,Ushioda-}. This is relevant to decay involving a neutrino in large-scale magnetized plasma; a detailed account will be presented in a future paper.
 
  The paper is organized as follows.   Section 2 presents features  of the electrons' classical motion, features and their quantum mechanical counter parts in a magnetic field.  Section 3 shows that a neutrino-photon interaction with  Chern-Simons form in a vacuum is not realized.  Section 4 derives an   anomalous  Chern-Simons form neutrino-photon interaction in magnetized plasma.     In Section 5, features of the anomalous neutrino photon interaction   are  studied.  Section 6 summarizes the study.   
\section{Uniform magnetic field  }
Charged particle trajectories   in  an external electromagnetic field differ greately   from vacuum ones. These make   quantum theories unique and different from those of no external fields. Hence, it is instructive to overview the motions of charged particles in an external electromagnetic fields  and its connection with quantum theories.
\subsection{Charged particles in a uniform magnetic field}

The classical motion of a charged particle a magnetic field is very different from that in no magnetic field.   \cite{taub} showed that all the trajectories are periodic for  ${\vec  B}^2> {\vec  E}^2$, where $\vec  E$ is the electric field.  A particle of a charge $q$ moves according to the Lorentz force $ q {\vec  v} \times {\vec  B}$, which is perpendicular to the velocity and does not change the kinetic energy of the particle. The motion of a charged particle is expressed by the Lagrangian $q {\vec  v}\! \cdot\! {\vec  A}$, and the Hamiltonian of such a particle is reduced to
\begin{eqnarray}
 H=\frac{( {\vec  p}+q {\vec  A})^2}{2m}, \label{Magnetic.Hamiltonian}
\end{eqnarray}
where $m$ is the mass and $(A_x,A_y)=\frac{B}{2} (-y,x)$ is a constant magnetic field in the $z$-direction. For an electron, $q=-e$, where $e$ is the unit of an electric charge, $|\vec  v|$ is at most the speed of light. The particle's motion conserves the energy  expressed by the Hamiltonian, which depends on the vector potential.  Consequently, the particle's trajectory depends implicitly on the vector potential. The particle energy also  increases as the vector potential increases in the region $x^2+y^2 \rightarrow \infty$. For example, the Lamor radius $R$ of circular motion is proportional to the velocity $v$, inversely proportional to the magnetic field, and given by $R=\frac{m}{eB}v$. The 
Hamiltonian becomes $H=\frac{(eBR)^2}{2m} $, which shows that the kinetic energy is proportional to $(BR)^2$. The particle energy in a constant magnetic field  becomes large in a large radius $R$. Furthermore, particle motion of a small $R$ in a large $B$ is equivalent to  that of a large $R$ in a  small $B$ as far as the product of $R$ and $B$ is constant. The effect of the magnetic field should appear in microscopic and macroscopic systems.      

 Quantum systems are described by the same Hamiltonian, and a quantum effect due to the magnetic field should also appear in a small system with a strong magnetic field and a large system with a weak magnetic field. The free Hamiltonian of a many electrons system is given by  
\begin{eqnarray}
 \int d^3x\,\psi^{\dagger}(\vec{x})\frac{( {\vec {p} }+q {\vec  A})^2}{2m} \psi(\vec {x})
\end{eqnarray}
with quantized electron fields, where the vector potential is the same as the one in Eq.$(\ref{Magnetic.Hamiltonian})$, and the effect of the magnetic field increases equally in    many-electron systems.

The space-time symmetry of this system of electrons in the uniform magnetic field differs from that in a vacuum. The vacuum is Lorentz invariant, and the fluctuations of the vacuum are also Lorentz  invariant. The magnetic field is not Lorentz invariant, but is given by the Lorentz tensor, $F_{\mu\nu}$, and varied  by the Lorentz transformation, $ \Lambda_{\mu\mu'}$ to $\tilde F_{\mu \nu}= \Lambda_{\mu\mu'} \Lambda_{\nu\nu'} F^{\mu'\nu'}$.  In a moving frame, a physical system of matter in an external magnetic field in $z$ direction, $F_{\mu \nu}=B, $ \text{for} $\mu =x, \nu =y; F_{\mu\nu}=0,\text{ for other} ~\mu, \nu $, has the tensor field $\tilde F_{\mu \nu}= \Lambda_{\mu\mu'} \Lambda_{\nu\nu'} F^{\mu'\nu'}$. Here, the transformed field, $\tilde F_{0i}$ is different from zero, and $ \tilde F_{\mu \nu} $ is different from $ F_{\mu \nu} $. The fluctuations of matter in the system of ${\vec  B} \neq 0, { \vec E=}0$ differs from those in the system of  $ { \vec  B} \neq 0, {\vec  E} \neq 0$. An effective Lagrangian of the system is not manifestly Lorentz invariant.  In this study, a matter Lagrangian in a system of uniform fields of ${\vec B} \neq 0, {\vec  E}=0$ that  is not invariant under vacuum fluctuations is obtained.
Quantum phenomena intrinsic to the magnetic field should appear.

Motion in the magnetic field is characterized by a radius $R$. The radius of the circular motion of an electron $R$ varies depending on the magnetic field and a particle's velocity. 
Figure 1 shows the radius $R$ for the magnetic field $B$ and velocity obtained from its kinetic energy, $\frac{m_e}{2} v^2 =\frac{3}{2} kT$ for the temperature $T=10^6\ \text{K}$ and for $T=300\ \text{K}$.     
\begin{figure}
\begin{center}
 \includegraphics[scale=.5]{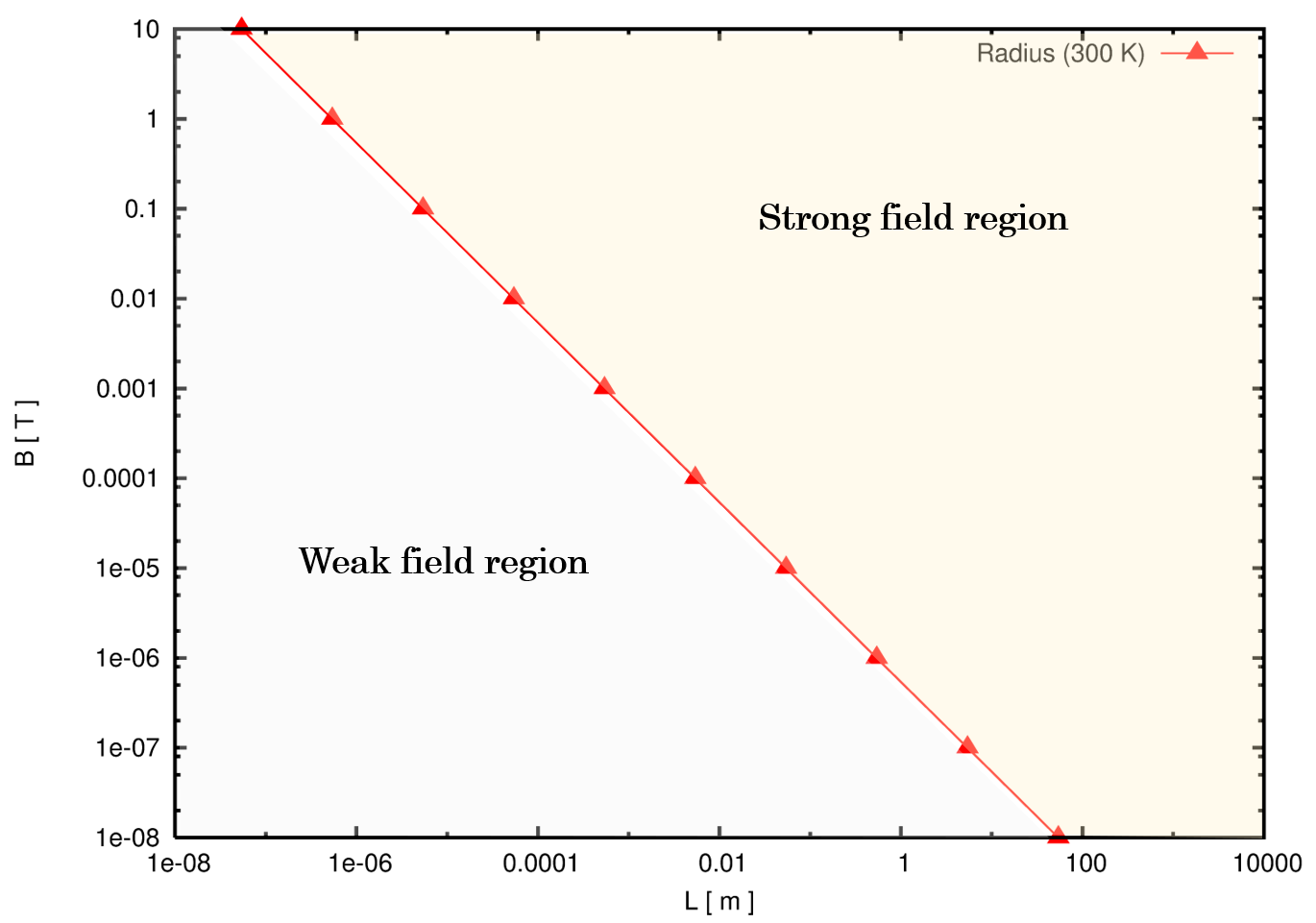}
\includegraphics[scale=.5]{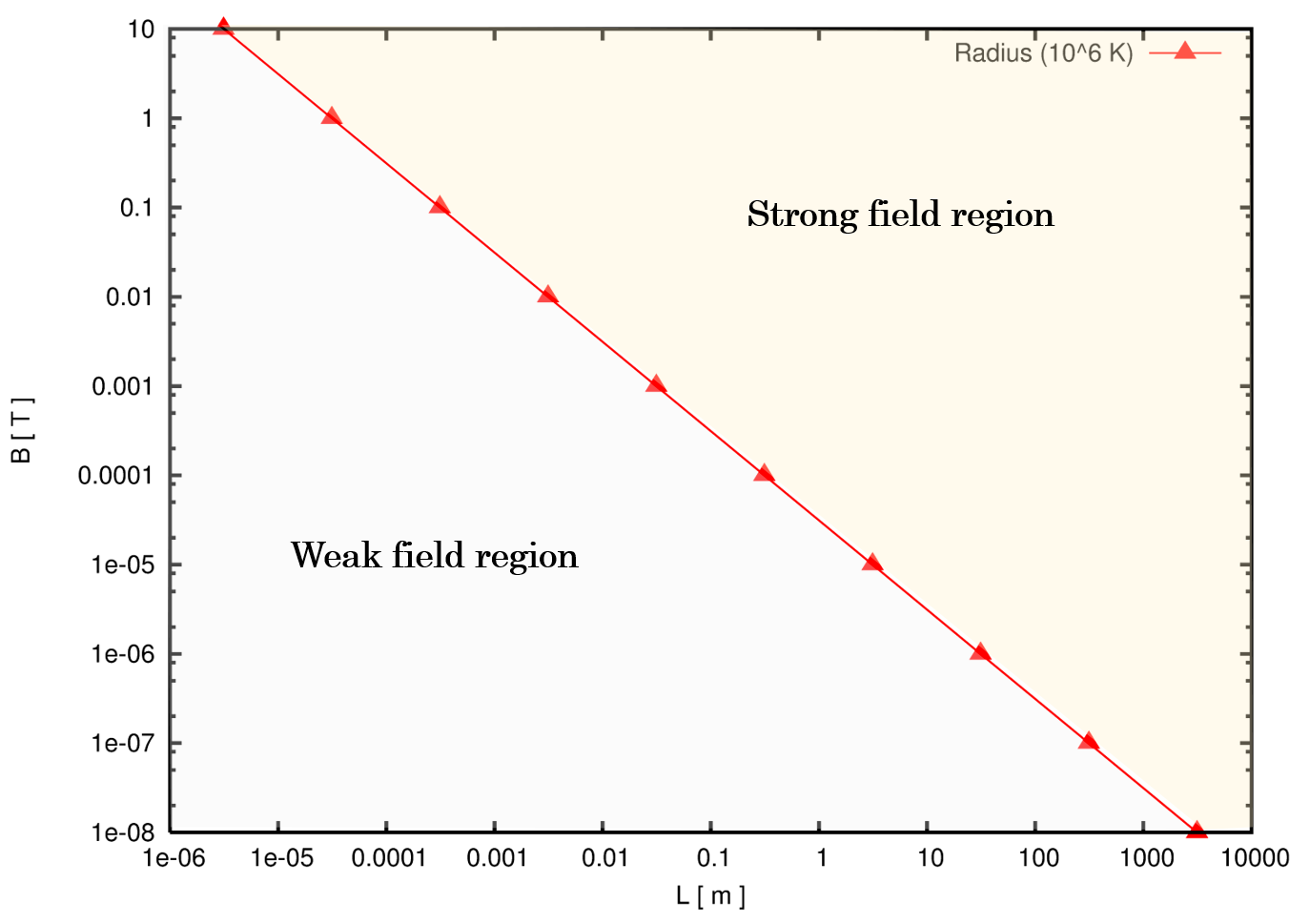} 
\caption{ The red line shows the Larmor radius  as a function of the magnetic field.  
The magnetic field and Larmor radius are in the vertical and holizontal axes, respectively.  
In the yellow region, the system is longer than the Larmor radius. The top  and bottom graphs   are at $300$ K and $10^6$ K, respectively.
}
\end{center}
\end{figure}
Quantum effects are important when the radius $R$ is smaller than the system's size, and are studied by the quantization of periodic orbits. 
Effects specific to the magnetic field appear in a large system for a weak magnetic field or large velocity.

The Larmor radius and mean free path of a GaAs-like semiconductor are presented in Fig. 2. 
The above condition is satisfied in the gray area. 
In Fig. 2, impurities and interactions are used in computing the mean free path
\footnote{The mean free path $l_{mf}$ is expressed with the collision time $\tau$ and velocity $v$ as $l_{mf}= \tau v=\frac{m v}{e} \times \mu$. 
The inverse of the mobility is the sum $\frac{1}{\mu}=\frac{1}{\mu_i}+\frac{1}{\mu_L}$, 
where $\mu_i \approx T^{\frac{3}{2}} $ and $\mu_L \approx T^{-\frac{3}{2}}$, where $T$ is the temperature
}.    
\begin{figure}[h]
\begin{center}
 \includegraphics[scale=.5]{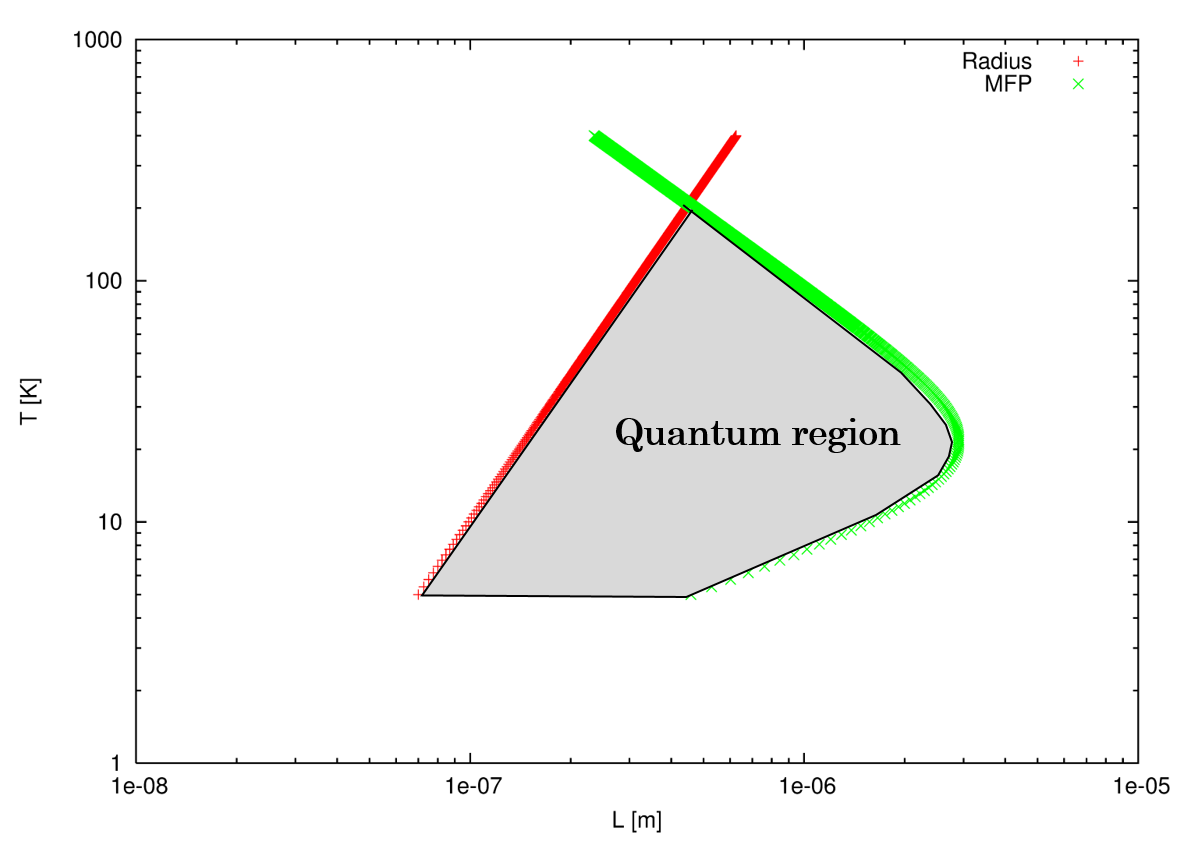}
\caption{
 The red line shows the Larmor radius of the velocity determined from the  temperature, and the green line shows the mean free path.  
The horizontal and vertical axes represent the length and temperature, respectively.      
In the gray region, the mean free path is larger than the Larmor radius for a GaAs-like semiconductor, and quantum effects are important.  
Temperature: $5 - 400$ K, Magnetic field: $1$ T. }
\end{center}
\end{figure}
The physical system is characterized by its size, particle energy, and uniform magnetic field.  
The system with a size and energy of $1$ mm and $1$meV, respectively, is equivalent to that with a size and energy of $10^9\ \text{mm}$ and $10^9\ \text{meV}$, respectively, at $10^9$ times its magnetic field. 
A phenomenon specific to a strong field appears even though the magnetic field is much lower than the critical field of $10^9$ T equivalent  to the rest energy of an electron $m_e c^2$. This effect may be rellevant to a large system of a uniform magnetic field.

\begin{figure}[h]
\begin{center}
 \includegraphics[scale=.5]{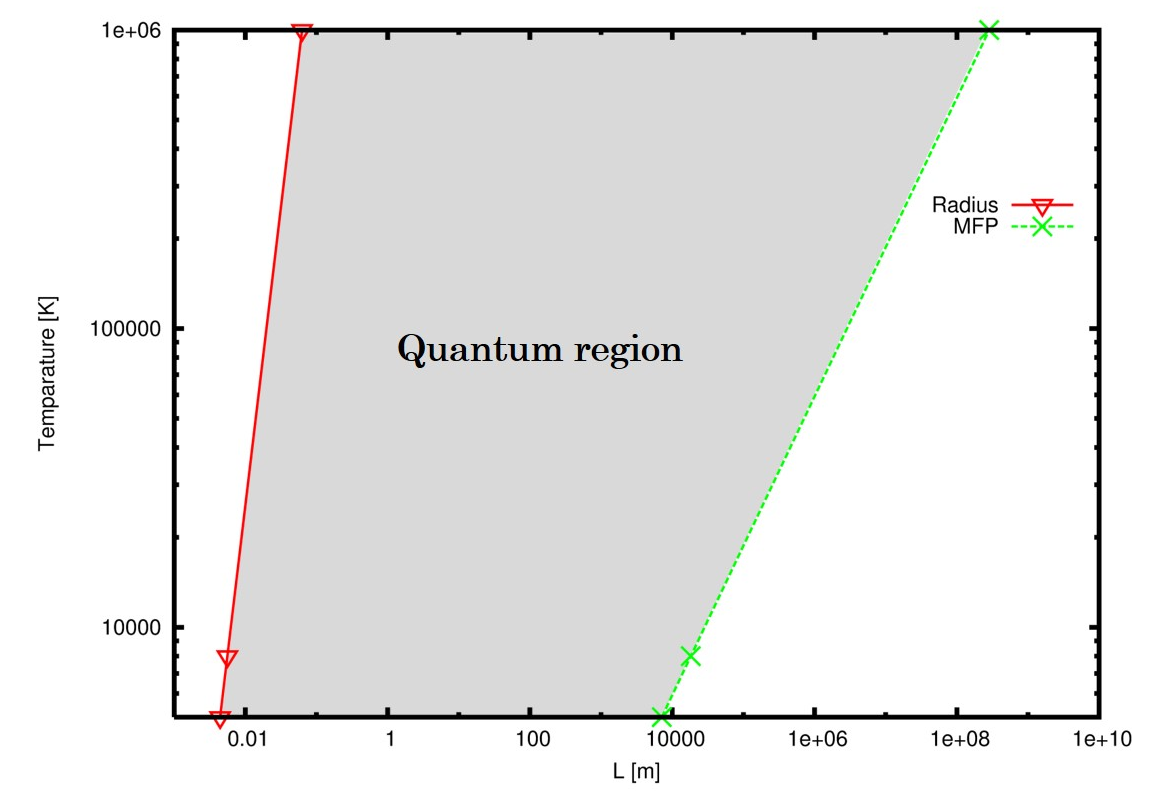}
\caption{
The red line shows the Larmor radius  of the velocity determined from the  temperature, and the green line shows the mean free path. The holizontal and vertical axises represent length and temperature, respectively.  In the gray region, the mean free path is larger than the Larmor radius for 
the solar corona parameters. B$=$0.0005 T, $N_e =10^{11}[\text{m}^{-3}]$. EW-Hall effects appear in the gray area of
$\text{Larmor radius} \leq \text{Scale of system} \leq \text{Mean free path}$.
}
\end{center}
\end{figure}

\subsection{Neutrino in a magnetized plasma}
 Higher order corrections due to electron fluctuations in a magnetized plasma reflect features of the magnetic field. 

 A system of electrons, photons, and neutrinos in a magnetic field has various unique properties.      

1. A magnetic field is not a Lorentz scalar, and a system of a constant magnetic field is not Lorentz invariant. Fluctuations of matter in a system of a constant magnetic field are described by an effective Lagrangian that is  not Lorentz invariant.  Accordingly, the  neutrino-photon interaction is different from the ones in a vacuum.      

2. The one-body Hamiltonian of a charged particle in a uniform magnetic field is equivalent to a harmonic oscillator, which has no energy upper bound. There is no threshold energy beyond which the effect of the magnetic field disappears, and a large system with a small magnetic field reveals the characteristics of a small system with a large magnetic field.  

3. Manifest Lorentz invariance is not satisfied. This leads to a neutrino decay probability that is not proportional to the invariant combinations of the neutrino momenta, and expressed by 
the Lorentz non-invariant combinations of the momenta. Details of this probability are given  
in Sec.(5.2.2.).


Electron fluctuations becomes prominent when a quantum coherence is fully preserved.  The mean free path gives the length that preserves the quantum coherence of an electron. The mean free path is inversely proportional to matter density, and large in dilute systems.  Long coherence is realized in dilute systems. 
Figure 3 shows a two-dimensional plot of the Larmor radius on the horizontal axis and  the temperature on the vertical axis in a magnetic field of $0.0005$ T and electron density of  $10^{11}$ ${\text m}^{-3}$. 
This corresponds to the parameters of the quiet region in the solar corona.  
The quantum region specified by the gray region exists.  

Landau levels express electrons in a uniform magnetic field \cite{Landau-level}. A field equivalently expresses the electron field of one component with many components. The electron propagator is classified by a topological invariance \cite{ishikawa-matsuyama} and reveals topological phenomena. 
The von-Neumann representation preserves these properties and is useful for finding an effective Lagrangian of the neutrino-photon system. This  Lagrangian is gauge invariant and  not manifestly Lorentz invariant. The induced interaction has  topological origins, intriguing  properties,  and a significant coupling strength proportional to $\frac{\rho_e}{B}$. 

The effective Lagrangian has been calculated mostly using an electron propagator in a  vacuum.

\section{ Neutrino-photon vertex of an Electroweak standard model for a vacuum}
 We study a system of electrons, photons, and neutrino for a finite electron density. The Lagrangian density describes the electroweak gauge theory of leptons: 
\begin{align}
 \mathcal{L}=\mathcal{L}_{EW}\bigl(\l(x),\nu(x),A_{\mu}(x), W_{\mu}(x),Z_{\mu}(x), \phi(x)\bigr) +\mathcal{L}_{gauge-fixing}(\xi) \label{electroweak}
\end{align}
where $\l(x)$, $\nu(x)$, $A_{\mu}(x)$, $W_{\mu}(x)$, $Z_{\mu}(x)$, and $\phi(x)$ are the lepton, neutrino, photon, and charged weak, neutral weak, and unphysical scalar bosons. The Higgs scalar is irrelevant and not discussed in this paper.   $\xi$ is the gauge parameter.
\subsection{ Weak $B$ expansion ($B^n, n \geq 0$ )  in the  $3+1$ dimensional field theory}
An induced interaction of the neutrino with  a photon caused by a loop correction of free electrons is applied to systems where electrons do not have closed paths. This section discuss the case of an  electron propagator in a vacuum, where higher order corrections for 
external field proportional to $B^0$ or to $B^n, n >0$ are derived.

Integrating the electron fields in the pure QED, of Eq. $(\ref{electroweak})$  in a system of external fields, an effective Lagrangian in a vacuum was first obtained by Schwinger with the proper time formulation as \cite{schwinger},
\begin{align}
\mathcal{L}= -\frac{1}{4}  F_{\mu\nu}F^{\mu\nu}  + \frac{2 \alpha^2}{45}\frac{1}{(m_e c^2)^4} \bigl[({\vec E}^2-{\vec  B}^2)^2+7(\vec  E\cdot \vec B)^2+\cdots\bigr]. \label{schwinger}
\end{align}
 This contribution is from the short distance fluctuation and manifestly Lorentz invariant and is the sum of positive powers of field strengths. This is equivalent to the one perturbatively obtained from the interaction Hamiltonian, $e^{-\epsilon t} H_{int}$.   
The effective interaction was obtained similarly in the electroweak theory:   
 \begin{align}
&\delta \mathcal{L}_{B} =-i  \mu_{ij} \bar \nu_i (x) \vec {\sigma}   \nu_j(x)\cdot \vec {B} \label{transition moment},\\
&\mu_{ij}=\frac{e G_F}{8{\sqrt 2} \pi^2 }(m_{\nu_i}+m_{\nu_j}) \sum_{l=e,\mu,\tau}f(a_l) U_{li}^{*} U_{lj},
 \end{align}
 where $a_l=\frac{m_l^2}{M_W^2} \ll1$, $f(a_l)\approx \frac{3}{2}(1-\frac{a_l}{2})$, and $U_{li}$ is the lepton mixing matrix  \cite{Raffelt, Fujikawa-Shrock}.  
The magnitudes are smaller than $10^{-23}\mu_B$, where $\mu_B=\frac{e\hbar }{2m_e}$.  
 
Vacuum fluctuations are due to an electron and positron pair. These have more energy than  twice the rest mass, $2m_e c^2$. 
Accordingly, the energy of the electromagnetic field in a vacuum, Eq.$(\ref{schwinger})$, and the energy of a neutrino in a magnetic field in a vacuum, Eq.$(\ref{transition moment})$, are proportional to a positive power of $\frac{eB}{m_ec^2}$. The contribution  of the short distance fluctuation in the magnetized plasma has the same properties, and  all terms 
of $B^n $, from $n=0$ to $n=\infty$, were obtained   
\cite{Battacharya}.   $\frac{1}{B}$ term does not appear in these cases.

\subsection{ Coupling strength of the electroweak Hall interaction in the electroweak gauge theory in the vacuum} 
 The effective Lagrangian in the vacuum or a system of a weak magnetic field is derived.
 
The Lagrangian density describes the electroweak gauge theory in an effective potential: 
\begin{eqnarray}
\mathcal{L}_{EW}(\l(x),\nu(x),A_{\mu}^{ext}+A_{\mu}, W_{\mu}(x), Z_{\mu}, \phi(x))+\mathcal{L}_{gauge-fixing}+\mu_e  e(x)^{\dagger} e(x) 
\end{eqnarray}
 $A_{\mu}^{ext}$ and $\mu_e$ are the external electromagnetic and chemical potentials of the electron, respectively. In the $R_{\xi}$ gauge \cite{fujikawa-lee-sanda}, the propagator of the weak boson in the momentum space, $W_{\mu \nu}(p)$, propagator of the unphysical scalar field in the momentum space, $H(p)$, and  electron propagator, $S(p)$, are  
\begin{align}
W_{\mu\nu}(p)=& -i\left(g_{\mu\nu}-\frac{p_{\mu}p_{\nu}}{M_W^2}\right)\frac{1} {p^2-M_W^2}  -i\frac{p_{\mu}p_{\nu}}{M_W^2}\frac{1}{p^2-M_W^2/\xi},\label{W_propagator}\\
 H(p) =&  \frac{i}{p^2-M_W^2/\xi}, \label{Higgs_propagator}\\
S(p) =& \frac{ p^\mu \gamma_\mu +m}{p^2-m^2}\label{electron_propagator},
\end{align} 
where 
$M_W$ is the mass of the W-boson. 
The interaction Lagrangians of the neutrino-lepton-$W_{\mu}$ and neutrino-lepton-$\phi$ are
\begin{align}
\mathcal{ L}_{e \nu W_{\mu}}(p) &= g \bar e(x)\Gamma_{\mu} \nu(x) W^{\mu}(x), \Gamma_{\mu}= \gamma_{\mu}(1-\gamma_5)/2, \\
\mathcal{ L}_{e \nu \phi}(p) &=\frac{m_e}{v_h} \bar e(x)\Gamma_{-}   \nu(x) \phi(x), \Gamma_{-}=(1-\gamma_5)/2,
\end{align}
where $g$ is the coupling strength and $v_h$ is the vacuum expectation of the Higgs field.
Their product agrees with the weak boson mass,  
\begin{eqnarray}
gv_h=M_W.
\end{eqnarray}

\begin{figure}[h]
\begin{center}
 \includegraphics[scale=.7]{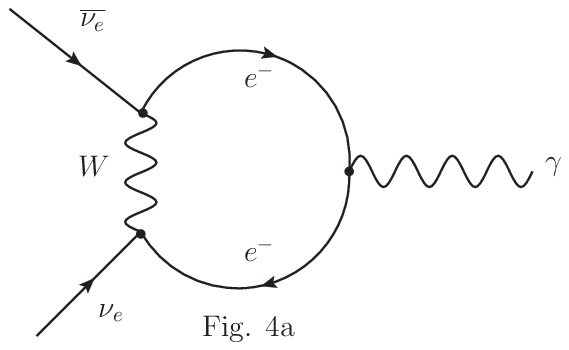}
\caption{
Feynman diagrams of Eq. $(\ref{W-loop})$  for $W_\mu$ (Fig. 4a).
}
\end{center}
\end{figure}

The Feynman diagram in Fig.4a shows the $\nu \nu \gamma$ amplitude in the coupling order of $g^2 e$. This is written with the weak boson and electron propagators in the coordinate spaces  $W^{\mu\nu}(x-y)$ and $S(x-y)$, respectively, as  
\begin{align}
\mathcal{ M}_1=g^2 e \int d^4x d^4y d^4z \bar \nu(x) \Gamma_{\mu} S(x-z) W^{\mu\nu}(x-y) \gamma^{\rho} S(z-y) \Gamma_{\nu} \nu(y) A_{\rho} (z), \label{W-loop} 
\end{align} 
where the Fourier transform gives the propagator in the coordinate space
\begin{align}
W^{\mu\nu}(x) &= \int \frac{d^4 p}{(2\pi)^4} e^{ipx} W^{\mu\nu}(p),  \label{Fourier_W}\\
H(x) &=\int \frac{d^4 p}{(2\pi)^4} e^{ipx} H(p), \label{Fourier_H}\\
S(x) &= \int \frac{d^4 p}{(2\pi)^4} e^{ipx} S(p)\label{Fourier_e}.
\end{align}
Another  diagram  of the same order is written with the scalar propagator in the coordinate space, $H(x-y)$, and  the electron propagator, $S(x-y)$, as 
\begin{align}
\mathcal{M}_2=\left( \frac{m_e}{v}\right)^2 e \int d^4x d^4y d^4z \bar \nu(x) \Gamma_{-}  S(x-z)  H(x-y) \gamma^{\rho} S(z-y) \Gamma_{-} \nu(y) A_{\rho} (z)\label{scalar-loop}.
\end{align} 
The sum of two diagrams is  expressed by $\tilde W^{\mu \nu}=W^{\mu \nu}(x-y)+ p^{\mu}p^{\nu} H (x-y)$
\begin{align}
\mathcal{M}&=\mathcal{M}_1+\mathcal{M}_2 \nonumber\\
&=g^2 e \int d^4x d^4y d^4z \bar \nu(x) \Gamma_{\mu} S(x-z) \tilde W^{\mu\nu}(x-y) \gamma^{\rho} S(z-y) \Gamma_{\nu} \nu(y) A_{\rho} (z), 
\end{align}
where 
\begin{align}
\tilde W^{\mu\nu}(x-y) =  \int \frac{d^4p}{(2 \pi)^4} e^{-ip(x-y)}\left[ -i\left(g_{\mu\nu}-\frac{p_{\mu}p_{\nu}}{M_W^2}\right)\frac{1} {p^2-M_W^2} \right] .
\end{align}
Using the short-ranged nature of $ \tilde W^{\mu\nu}(x-y)$, integrating over $x-y$ for a well-behaved function $F(y)$ can be expressed by the following series:     
\begin{align}
&g^2\int d^4(x-y) \tilde W^{\mu\nu}(x-y) F(y) \nonumber \\
&=g^2\int d^4(x-y) \tilde W^{\mu\nu}(x-y) \sum_{l} \frac{1}{l!}  (x-y)^{\mu_1} \cdots (x-y)^{\mu_l}\frac{\partial}{\partial x^{\mu_1}}\cdots \frac{\partial}{\partial x^{\mu_l}} F (x)\nonumber \\
&=g^2 C_0 (-i)g^{\mu\nu} \left(1+O\left(\frac{\partial_x }{M_W}\right)\right) F(x),\\
&C_0 g^{\mu\nu}= \int d^4(x-y) \tilde W^{\mu\nu}(x-y)=- \frac{g^{\mu\nu}}{M_W^2},
\end{align}  
where the higher power terms are proportional to $\frac{1}{M_W^n}, n\geq 2$ and neglected.
Substituting $F(y)=S(z-y) \Gamma_{\nu} \nu(y)$ and Taylor expanding the functions of variables $x$ and $y$,   
we have
\begin{align}
& \int d^4y  \tilde W^{\mu\nu}(x-y) \gamma^{\rho} S(z-y) \Gamma_{\nu} \nu(y)\nonumber \\
&=\int d^4y  \tilde W^{\mu\nu}(x-y) \left( 1+(x-y) \frac{\partial}{\partial x} + \cdots \right)\gamma^{\rho} S(z-x) \Gamma_{\nu} \nu(x) \nonumber \\
&=\left(-iC_0+\frac{1}{M_W} \frac{\partial}{\partial x} + \cdots \right)\gamma^{\rho} S(z-x) \Gamma^{\mu} \nu(x),
\end{align}
where $\cdots$ are proportional to $\frac{1}{M_W^n}; n \geq 2$ and negligible. Next, we express the vector potential $A_{\rho}(z)$ in the Taylor series of $z-x$ around $x$. The integral is written as
\begin{align}
& \int d^4x d^4y d^4z \bar \nu(x) \Gamma_{\mu} S(x-z) \tilde W^{\mu\nu}(x-y) \gamma^{\rho} S(z-y) \Gamma_{\nu} \nu(y) A_{\rho} (z)\nonumber \\
=& \int d^4x  d^4z \bar \nu(x) \Gamma_{\mu} S(x-z)   \gamma^{\rho} S(z-x)  \Gamma^{\mu}\nu(x)  (-iC_0)\left(1+O(\frac{1}{M_W}) \right) \nonumber \\
&\times \left(1+(z-x)_{\mu} \frac{\partial}{\partial x_{\mu}}+\text{higher~ derivative~ terms} \right) A_{\rho} (x)\nonumber\\
=&\int d^4x  \left[ \bar \nu(x) {I_1}^{\rho}    \nu(x)  A_{\rho} (x)+ \bar \nu(x) {I_2}^{\mu}_{\alpha}  \nu(x)  \frac{\partial}{\partial x_{\alpha}} A_{\rho} (x)\right] (-iC_0), \\
&{I_1}^{\rho}=\int  d^4z  \Gamma_{\mu} S(x-z)   \gamma^{\rho} S(z-x)  \Gamma^{\mu}, \\
&{I_2}_{\alpha}^{\rho} =\int  d^4z  \Gamma_{\mu} S(x-z)   \gamma^{\rho} S(z-x)  \Gamma^{\mu} (z-x)_{\alpha},
\end{align}
where $O(\frac{1}{M_W})$ and higher derivative terms were dropped.  We have
\begin{align}
&{\mathcal M}/{\frac{e G_F}{\sqrt 2}}=i \int d^4x  \left[ \bar \nu(x) {I_1}^{\rho}    \nu(x)  A_{\rho} (x)+ \bar \nu(x) {I_2}^{\mu}_{\alpha}  \nu(x)  \frac{\partial}{\partial x_{\alpha}} A_{\rho} (x)\right], 
\end{align} 
where 
\begin{eqnarray}
G_F=\frac{g^2}{M_W^2}.
\end{eqnarray}
\subsection{Charge and dipole moment of the neutrino}
$I_1^{\rho}$ and $I_2^{\rho}$ represent induced charge and show dipole moment, respectively. The charge should vanish as the neutrino beomes neutral; however, the dipole moment can be finite.
  
We evaluated $I_1^{\rho}$ first. Substituting the electron propagator, $(\ref{electron_propagator})$, we have   
\begin{align}
 I_1^{\rho} &={\gamma}_{\mu}(1-\gamma_5)  \int  d^4z   S(x-z)   \gamma^{\rho} S(z-x)  {\gamma}^{\mu} (1-\gamma_5) \nonumber \\
&=(1-\gamma_5) 4m \int \frac{d^4p}{(2\pi)^4}\frac{p^{\rho}}{(p^2+m^2)^2}  \nonumber \\
&=0.
\end{align}

Next, we evaluate $I_2^{\rho}$.   Using the identity 
\begin{eqnarray}
\gamma_{\mu}=\frac{\partial}{\partial p_{\mu}} S(p)^{-1},
\end{eqnarray}
and Eqs.$(\ref{W_propagator})$, $(\ref{Higgs_propagator})$, and $(\ref{electron_propagator})$,  we have   
\begin{align}
 {I_2}_{\alpha}^{\rho} &=\int  d^4z  \Gamma_{\mu} S(x-z)   \gamma^{\rho} S(z-x)  \Gamma^{\mu} (z-x)_{\alpha} \nonumber \\
&=\int  d^4z  \Gamma_{\mu} \int \frac{dp_1}{(2\pi)^4} e^{i p_1(x-z)} S(p_1)    \gamma^{\rho} \int \frac{dp_2}{(2\pi)^4}e^{ip_2(z-x)}  S(p_2)  \gamma^{\mu} \Gamma_{-}(z-x)_{\alpha} \nonumber \\
&=\Gamma_{\mu} \int \frac{d^4p}{(2\pi)^4}   \left(\frac{\partial}{\partial p_{\alpha}} S(p) \right) (-) \left(\frac{\partial }{\partial p_{\rho}}  S(p)^{-1}\right) S(p) \left(\frac{\partial}{\partial p_{\mu}} S(p)^{-1} \right)\Gamma_{-}  \nonumber \\
&=\Gamma_{\mu} \int \frac{d^4p}{(2\pi)^4} \left( S(p)  \frac{\partial}{\partial p_{\alpha}} S(p)^{-1}  S(p) \right) \left(\frac{\partial }{\partial p_{\rho}}  S(p)^{-1} \right) S(p) \left( \frac{\partial}{\partial p_{\mu}} S(p)^{-1}\right) \Gamma_{-}  \nonumber \\
&=\Gamma_{\mu} \int \frac{d^4p}{(2\pi)^4}  S(p) \left( \frac{\partial}{\partial p_{\rho}} S(p)^{-1} \right) S(p) \left(\frac{\partial }{\partial p_{\alpha}}  S(p)^{-1} \right) S(p)  \left( \frac{\partial}{\partial p_{\mu}} S(p)^{-1}\right)  \Gamma_{-} \nonumber \\
&=\frac{\Gamma_{\mu}}{4} \int \frac{d^4p}{(2\pi)^4} \text{Tr}\left[ S(p)  \left(\frac{\partial}{\partial p_{\alpha}} S(p)^{-1} \right) S(p) \left(\frac{\partial }{\partial p_{\rho}}  S(p)^{-1}\right)  S(p) \left( \frac{\partial}{\partial p_{\mu}} S(p)^{-1} \right) \right] \Gamma_{-} + \text{ others}  \nonumber \\
&=\frac{\Gamma_{\mu}}{2} \int \frac{d^4p}{(2\pi)^4} \text{Tr}\left[ S(p) \left( \frac{\partial}{\partial p_{\alpha}} S(p)^{-1} \right) S(p) \left(\frac{\partial }{\partial p_{\rho}}  S(p)^{-1}\right) S(p)  \left(\frac{\partial}{\partial p_{\mu}} S(p)^{-1}\right) \right]+ \text{ others}  \label{topological},   
\end{align}
where the identity $\Gamma_{\mu} \Gamma_-=\Gamma_{\mu}$ was substituted, a topological invariant term is extracted, and  terms proportional to other $\gamma$-matrices are not relevant and dropped.

$I_1$ vanishes, and a charge is not induced. The electromagnetic  gauge invariance guanrantees this condition. The dipole moment proportional to $I_2$ is manifestly gauge invariant and does not necessarily vanish. Equation $( \ref{topological} )$ shows that this is expressed by a topological invariant of the propagator. This invariance is finite and Lorentz invariant in $2+1$ dimensional Dirac theory but  vanishes in the $3+1$ dimensional Dirac theory. The system of the uniform magnetic field is studied next.        

We should note that the above result agrees with that derived from   Lagrangian density 
\begin{align}
&\mathcal L=   \mathcal L_0 +\frac{G_F} { \sqrt 2} \bar {e}(x)\Gamma_{\mu} e(x) \bar
 {\nu_e}(x)\Gamma_{\mu}{\nu_e}(x) +e j_{\mu}( A^{\mu}_{ext}+A^{\mu}), \\
& \mathcal L_0= \sum_i \bar {\nu_i}(x)(p^{\mu}\Gamma_{\mu}-m_{\nu_i}) {\nu_i}(x)+
\bar e(x)(p^{\mu}\gamma_{\mu}-m_e)e(x)-\frac{1}{ 4}F_{\mu\nu}F^{\mu\nu},
\label{F_lagrangian}
\end{align}
where $p^{\mu}=-i\frac{\partial}{\partial x_{\mu}}$. 
$A^{\mu}(x)$ and $A^{\mu}_{ext}(x)$ are the vector
potential and that of expressing the external magnetic field and the chemical potential.  The
neutral current interaction is symmetric in all neutrino flavors and  does not contribute to the neutrino radiative transitions and was ignored in
Eq. $(\ref{F_lagrangian})$. 

\section{ Neutrino-photon vertex in finite electrons density in the uniform magnetic field}
 This section discusses a many-body system of electrons in closed orbits realized in the yellow region in Fig. 1, and the quantum regions in Figs. 2 and 3. An electron propagator describes this system in the magnetic field. The electrons are in closed orbits, and their fluctuations 
are intrinsically different from the free electrons studied in the previous section. An effective Lagrangian is expressed by the topological invariant expression Eq. $( \ref{topological} )$ using the electron propagator $ S_{B}(p)$, in the magnetic field system. The topological invariant in the magnetic field does not vanish.     

\subsection{Symmetry }
\subsubsection{Electromagnetic gauge invariance }
 Gauge invariance holds in a quantum system with a magnetic field; however, inversion symmetry  is partly broken. 
 The Lagrangian  
\begin{eqnarray}
\mathcal{L}=\bar e (x)  \gamma^{\mu}\left(p_{\mu}+ e \left({A^{ext}}_{\mu}(x)+A_{\mu}(x)\right)\right) e(x)+ \cdots,
\end{eqnarray}
is gauge invariant 
\begin{align}
A_{\mu}(x) &\rightarrow U^{-1}(x)(A_{\mu}+p_{\mu} ) U(x), \\
 e(x) &\rightarrow U(x) e(x),
\end{align}
where $U=e^{i\lambda(x)}$. 
Electromagnetic current $J_{\mu}(x)=\bar e(x) \gamma_{\mu} e(x)$ satisfies  
\begin{eqnarray}
\partial^{\mu} J_{\mu}(x)=0,\ [J_{0}(x), e(y)]\delta(x_0-y_0)=- e (x) \delta^4(x-y),
\end{eqnarray} 
and the vertex part is defined by using momenta in the magnetic Brillouin zone defined later in Eq. $( \ref{magnetic Brillouin zone})$,
\begin{align}
\int d^4z d^4x d^4y e^{i(qz -p_1x -p_2y)} \langle 0|T(J_{\mu}(z) \bar e(x) e(y)) |0 \rangle\nonumber\\
=(2 \pi)^4 \delta_{MB}^{(4)}( p_1-p_2-q) S_{B}(p_1) \Gamma_{\mu} S_B(p_2),
\end{align}
where $MB$ represents the magnetic Brillouin zone.
The propagator is written with the Landau level basis, 
\begin{eqnarray}
S_B(x-y)=\int d \lambda \frac{1}{\lambda} \phi_{\lambda}(x) \phi_{\lambda}^{*}(y),
\end{eqnarray}
where $\lambda$ denotes $p_0$, $ p_z$, $ {\bs p}_{BZ} $, and one particle states are expressed with the Landau level wave functions $u_l({\bs p}_{BZ},x,y)  $ as
\begin{eqnarray}
\phi_{\lambda}(x)=e^{-i(p_0 x_0-p_z z)}u_l({\bs p}_{BZ},x,y). 
\end{eqnarray}
Ward-Takahashi identity holds in the system of the external magnetic field. The details are given in the following  section. 

 Under space inversion, $ \vec  x \rightarrow - \vec  x$, the vector potential is transformed as 
 \begin{align}
\vec A(x) \rightarrow - \vec  A(x),\ \nabla \times \vec  A \rightarrow \nabla \times \vec  A. 
\end{align}
Accordingly, the magnetic field is invariant:
\begin{align}
\vec  B \rightarrow \vec  B. \label{magnetic_field_P}
\end{align} 

The topologically invariant Eq.$( \ref{topological} )$ of the dipole moment of the neutrino with  the electron propagator $ S_{B}(p)$ in the magnetic field.
 
The dipole moment of the neutrino is expressed by the topologically invariant Eq.$( \ref{topological} )$ for an electron propagator $ S_{B}(p)$ in a magnetic field. The expression in the magnetic field is equivalent to that  in a vacuum; however the topological invariance does not vanish.   

\subsubsection{ Product of the currents  } 
 A transition amplitude of a neutrino with a photon is written as a product of the photon field with electron or neutrino currents  
from Eq.$(\ref{F_lagrangian}) $,
\begin{align}
\mathcal{M}=\int d^4x d^4y J_{\mu}({\nu}, x) \langle 0| T\left((J_{V}^{\mu}(x)-J_{A}^{\mu}(x)) \times J_{V}^{\nu}(y) \right)|0 \rangle A_{\nu}(y), 
\end{align}
where $A_{\nu}(y)$ is the photon field, and
\begin{align}
J_{\mu}({\nu},x) &= \bar \nu(x) \Gamma_{\mu} \nu(x),\\
J_V^{\mu}(x) &= \bar e(x) \gamma^{\mu} e(x),\ J_A^{\mu}(x)=\bar e(x) \gamma_5 \gamma^{\mu} e(x).
\end{align}
The magnetic field is invariant under the space inversion of Eq.$(\ref{magnetic_field_P})$ and this amplitude is reduced further to
\begin{align}
\mathcal{M}&=\int d^4x d^4y J_{\mu}({\nu},x) \langle 0|T\Bigl( \bigl(J_{V}(x)^{\mu}-J_{A}^{\mu}(x)\bigr) \times  J_{V}^{\nu}(y) \Bigr) |0 \rangle  A_{\nu}(y) \nonumber \\
&= \int d^4x d^4y J_{\mu}({\nu}, x) \left( \pi_{VV}^{\mu \nu}(x-y)-\pi_{VA}^{\mu\nu}(x-y)\right)  A_{\nu}(y) \nonumber \\
&=\int d^4x d^4y J_{\mu}({\nu},x)  \pi_{VV}^{\mu \nu}(x-y)  A_{\nu}(y),
\end{align}
where 
\begin{align}
\pi_{VV}^{\mu \nu}(x-y)&=\langle 0|T( J_V^{\mu} (x)\times   J_V^{\nu}(y) ) |0 \rangle,  \\
\pi_{AV}^{\mu \nu}(x-y)&=\langle 0|T( J_{A}^{\mu}(x)\times J_{V}^{\nu}(y) ) |0 \rangle,
\end{align}
where  $ \pi_{AV}^{\mu \nu}(x-y)= 0$ due to parity invariance of the electron loop diagram. The amplitude is given by the current correlation function of the vector current.

 \subsection{Strong ${B}$ expansion in the system of finite electron density}     
We compute an effective Lagrangian and obtain the dipole moment of a neutrino using  the Ward-Takahashi identity for electrons in a magnetic field. The dipole moment reveals $\frac{1}{B}$ behavior.  

A momentum representation for  electrons  in the Landau levels is suitable for studying an electromagnetic interaction of a neutrino induced by  many-body quantum effects. 
The von Neumann representation has this property and is applied in the present work.

The system of finite electron density and a magnetic field parallel to the $z$-axis is studied. 
The magnetic field and chemical potential $\mu_e$ are described by  $(A^{ext}_x,A^{ext}_y)=\frac{B}{2}(-y,x)$ and  $A^{{ext}}_{0}(x)=\frac{\mu_e}{e}$. 
In this system, the electrons of the negative energies and the energy $0 \leq E \leq \mu_e$ are occupied. The effective Lagrangian from the former is that of the vacuum and was obtained in \cite{erdas}. The latter  is studied in the present paper.  

\subsubsection{von Neumann lattice representation}

The electrons occupy the energy level $0 \leq E \leq \mu_e$, and fluctuations have no energy gap. Moreover, the fluctuation extends spatially up to infinity unless forbidden 
by boundary conditions. The classical circular motion of the radius proportional to $\frac{1}{B}$ may result in a $\frac{1}{B}$ term in the infinite system. 
The effective Lagrangian may possess a long-distance component due to sizeable circular motion.  
This may break Lorentz invariance and have a different dependence on the magnetic field  from that in a vacuum.  
We focus on the effect of these non-relativistic electrons. 

In the following, we present the case of low-energy electrons for simplicity.  
The  high-energy region can be straightforwardly covered using relativistic Landau levels, which are given in the appendix.   
The single electron Hamiltonian is
\begin{align}
{H}_0=  \frac{({p_{x}}+e{ A}_x^{ext})^2}{  2m}+\frac{({ p_{y}}+e{ A}_y^{ext})^2}{  2m}+\frac{p_z^2}{  2m}+\hbar \sigma_z \frac{eB}{2m}. 
\end{align}
The last term on the right-hand side is for an electron with a magnetic moment. The system is invariant under space inversion $(x,y) \rightarrow -(x,y)$ as the external vector potential transforms $\vec A^{ext} \rightarrow -\vec A^{ext} $ but is not invariant under time inversion $t \rightarrow -t$ owing to the magnetic field $B$.  
To elucidate the interactions of the neutrino with photons, we decompose the electron field in terms of a free wave in the $z$-direction and those  of the Landau levels in the 
$(x,y)$ space.  

Because the electron's motion is circular around an arbitrary center, the electron field is expressed  with center and relative coordinates $(X,Y)$ and $(\xi,\eta)$, respectively. 
A one-body Hamiltonian in the $(x,y)$ space is described by the relative coordinates $(\xi,\eta)$, and commutes with center coordinates $(X,Y)$,   
\begin{align}
&\xi =(eB)^{-1}(p_x+eA_x),\ \eta =(eB)^{-1}(p_y+eA_y),\\
& X=x-\xi,\ Y=y- \eta, \\
&[ \xi,\eta ]=-[X,Y]=\frac{a^2}{2 \pi i},\ a=\sqrt{\frac{2 \pi \hbar}{eB}},\\
&{H^0}_{(xy)} = \frac{1}{2} m {\omega_c}^2 (\xi^2+\eta^2)+\hbar \frac{\sigma_z}{2}\omega_c,\ \omega_c=\frac{eB}{m}.
\end{align}
The eigenstates of this Hamiltonian and those of the center coordinates,  
\begin{align}
{H^0}_{(xy)}|l \rangle &= E_l| l \rangle,\ E_l= (l+\frac{1}{2} \pm\frac{1}{2}) \hbar \omega_c,\   l=0,1,2, \cdots,  \\
(X+ iY) |\alpha_{mn} \rangle &=a z_{mn} | \alpha_{mn} \rangle,\ z_{mn}= (m \omega_x+i n {\omega_y} ), \\
|\alpha_{mn} \rangle &= e^{i \pi(m+n+mn)+\sqrt{\pi} (A^{\dagger}\frac{z_{mn}}{a}-A\frac{z_{mn}^{*}}{a})} |\alpha_{00} \rangle,\\
 A &=\frac{\sqrt \pi}{a}(X+iY),\ [A,A^{\dagger}]=1,
\end{align}
where $\omega_x$ and $\omega_y$ are complex numbers satisfying 
\begin{eqnarray}
\text{Im} [\omega_x^{*} \omega_y]=1.
\end{eqnarray}
As the vector potential becomes infinite at $\sqrt{x^2+y^2} \rightarrow \infty$, all the states in $(x,y)$ space are bound states. The quantum effect of the magnetic field is not cut in the short and long-distance regions and appears even in high-energy regions. The many-body effects due to their fluctuations are studied easily with the  momentum states of the guiding centers:
\begin{align}
|\alpha_{\bs p} \rangle &= \sum_{mn}e^{ip_x m+ip_y n} |\alpha_{mn} \rangle , \langle \alpha_{\bs p}|\alpha_{\bs p'} \rangle=\alpha({\bs p}) (2\pi)^2 \delta({\bs p}-{\bs p}'-2 \pi  \bs N),\\
| \beta_{\bs p} \rangle &= \frac{1}{\beta(\bs p)} |\alpha_{\bs p} \rangle, \alpha( \bs p)= \beta^{*}( \bs p)  \beta( \bs p),
\end{align}
where $\bs N$ is a two-dimensional vector of the integer components, $\bs p$ is the two-dimensional momentum in the magnetic Brillouin zone, and $\alpha(\bs p)$ is expressed by the Jacobi $\theta $ -function. The von Neumann basis is convenient to investigate the many-body dynamics of the electrons in the external magnetic
field \cite{imos,iaim}. 

The $z$-direction is treated with the plane wave of the momentum eigenstate  $|p_z \rangle$. The spin freedom is decoupled from the momentum  and 
is ignored in the following discussions.    
Expanding the electron field with the basis as 
\begin{align}
e({\vec x},t)&=\int\frac{d p_z}{ 2\pi} \int_{BZ}\frac{d^2 p}{ 
 (2\pi)^2}\sum_{l=0}^{\infty} b_l({\bs p},p_z,t) \langle {\bs
 x}|l,\beta_{\bs p},p_z \rangle,\\
 E_l(p_z)&=E_l+\frac{p_z^2}{ 2m},\  
\langle {\vec x}| l,\beta_{\bs p} , p_z \rangle = \langle {\bs x}| l,\beta_{\bs p}  \rangle \cdot \langle z|p_z \rangle,
\end{align}
where $E_l$ is the energy of the Landau level and the BZ represents for the magnetic Brillouin zone  
\begin{eqnarray}
p_x, p_y \leq \frac{\pi}{ a}. \label{magnetic Brillouin zone}
\end{eqnarray} 
The action including the fluctuations of the effective electromagnetic field 
 $\tilde A_{\mu}=e A_{\mu}+\frac{G_F}{  \sqrt 2}\bar
 \nu(x)\gamma_{\mu}(1-\gamma_5) \nu(x)$ is  
\begin{align}
S &=S_0+ \int d^4x \left[  j_0(x) {\tilde A}_0(x) - {\vec j}(x)\cdot{\vec {\tilde A}}(x) +O({\tilde A}_{\mu}^2)\right] \label{action} ,\\
S_0 &= \int dt \frac{d p_z}{ 2\pi} \int_{BZ}\frac{d^2 p }{  (2\pi)^2}
 b_l^{\dagger}({\boldsymbol p},p_z\,) 
\left[i\hbar \frac{\partial}{
 \partial t} -E_l(p_z)
\right]
 b_l ({\boldsymbol  p},p_z),
\end{align}
where the current operators $j_0(x)$ and ${\vec j}(x)$ are given below. 
The creation and annihilation operators satisfy
\begin{align}
&\bigl\{ b_l({\bs  p},p_z\,), b_{l'}^{\dagger}({\bs p}',p_z') \bigr\}= \delta_{l l'}(2
 \pi)\delta(p_z-p_z')(2 \pi)^2 ({\bs p}-{\bs p}\,'-2\pi{\bs N})
 e^{i\phi({\bs p},{\bs N})},\\
&\phi({\bs p},{\bs N})= \pi(N_x+N_y)-N_yp_x,\ N_i: \text {integer}.
\end{align}
The one-body Hamiltonian is diagonal in the action Eq.$(\ref{action})$: however the density and current in the $(t,x,y,z)$ space  are expressed as
\begin{align}
j_0({\bs k},k_z)&= \int d^3 x  e^{-i({\bs k} {\bs x}+k_z z) } e^{\dagger} (x) e(x) \nonumber \\
 &=\sum_{l,l'} \int \frac{dp_z}{ 2\pi} \int_{BZ}\frac{d^2p}{
 (2\pi)^2} { b}_l^{\dagger}({\bs p},p_z) \Gamma_0({\bs k})_{l,l'} b_{l'}({\bs p}+a\hat k,p_z+k_z)   \label{form-factor_1},\\
{\bs j}({\bs  k},k_z)&=\int d^3 x  e^{-i({\bs k} {\bs  x} +k_z z)} e^{\dagger} (x)  {\bs v} e(x) \nonumber \\
 &= \sum_{l,l'} \int \frac{dp_z}{ 2\pi} \int_{BZ}\frac{d^2p}{
 (2\pi)^2}  b_l^{\dagger}({\bs p},p_z){\bs \Gamma({\bs k})}_{l,l'} b_{l'}({\bs p}+a\hat k,p_z+k_z) \label{form-factor_2}, \\ 
{ j}_z({\bs  k},k_z)&=\int d^3 x  e^{-i({\bs k} {\bs  x} +k_z z)} e^{\dagger} (x)   v_z e(x) \nonumber \\
 &= \sum_{l,l'} \int \frac{dp_z}{ 2\pi} \int_{BZ}\frac{d^2p}{
 (2\pi)^2}  b_l^{\dagger}({\bs p},p_z) \frac{p_z}{m}\Gamma_0({\bs k})_{l,l'} b_{l'}({\bs p}+a\hat k,p_z+k_z),
\end{align}
where
\begin{align}
\bs v &=\frac{\bs p+e\bs A_{ext}}{m},\ v_z=\frac{p_z}{m}, \\
\hat k_i &=W_{ij} k_j,\\
W&=
\left(
\begin{array}{cc}
\text{Re} [ {\omega}_x] & \text{Im} [ {\omega}_x]  \\
\text{Re} [ {\omega}_y] & \text{Im} [ {\omega}_y]
\end{array}
\right) 
\end{align}
In the above equations, $\bs \zeta=(\xi,\eta)$ and $\hat {\bs k}$ are  the relative
coordinate and dual momentum of ${\bs k}$, and
\begin{align}
\Gamma_0({\bs k})_{l,l'} =\langle f_l| e^{-i{\bs k}\cdot{\bs 
 \zeta}}  | f_{l'}  \rangle,\  
{\bs \Gamma({\bs k})}_{l,l'}= \langle f_l| \frac{1}{ 2}\{{\bs v}, e^{-i{\bs k}\cdot{\bs
 \zeta}} \} | f_{l'} \rangle. 
 \end{align}
The form factor $\Gamma_0(\bs k)$  in
 Eq.$(\ref{form-factor_1} )$ and $\bs \Gamma(\bs k)$  in
 Eq.$(\ref{form-factor_2} )$ is expressed by  polynomials in $\bs k$. 
The constant term, $\Gamma_0(0)$, is proportional to the electron charge, and the linear term $\frac{\partial }{\partial k_i} \Gamma(0)$ is proportional to the dipole moment, which has non-vanishing matrix element  between the nearest neighbor Landau levels, that is, $l'=l \pm 1$.

Transforming electron operators as
\begin{eqnarray}
\hat b_l(\bs p)=\sum_{l'}U_{l l'}(\bs p) b_{l'}(\bs p)
\end{eqnarray}
with the unitary matrix
\begin{eqnarray}
U_{l l'}(\bs p)=\langle f_l | e^{i{\bs p} \cdot {\bs {\hat  \xi}}/a-\frac{ i}{4\pi} p_x p_y}|f_{l'} \rangle,\ \hat \xi_i=W_{ij} \xi_j,
\end{eqnarray}   
is conveinient. These operators satisfy  
 \begin{align}
&\bigl\{ {\hat b}_l({\bs p},p_z\,), {\hat b}_{l'}^{\dagger}({\bs  p}',p_z') \bigr\}= \delta_{l l'}
(2\pi)^3\delta(p_z-p_z')({\bs p}-{\bs p}\,'-2\pi{\bs N}) \Lambda_{l l'}(\bs N),\\
&{\hat b}_l({\bs p}+2 \pi {\bs N},p_z)=  \Lambda_{l l'}(\bs N) {\hat b}_{l'}({\bs p}, p_z),\\
& \Lambda_{l l'}(\bs N)=(-1)^{N_x+N_y} U_{l l'}(2 \pi \bs N),
 \end{align}
and lead   the simple expression of  the charge and current operators 
 \begin{align}
&j_0({\bs k},k_z)= \sum_{l} \int \frac{dp_z}{ 2\pi} \int_{BZ}\frac{d^2p}{
 (2\pi)^2} {\hat b}_l^{\dagger}({\bs p},p_z)  {\hat b}_{l}({\bs p}+a\hat k,p_z+k_z)   \label{form-factor_{20}},\\
&{\bs j}({\bs k},k_z)= \sum_{l,l'} \int \frac{dp_z}{ 2\pi} \int_{BZ}\frac{d^2p}{
 (2\pi)^2}  {\hat b}_l^{\dagger}({\bs p},p_z) \left\{ \bs v+\frac{a \omega_c}{2\pi}(W^{-1}\bs p+\frac{a}{2} \bs k)\right\}_{l,l'} {\hat b}_{l'}({\bs  p}+a\hat k),\\
&j_z({\bs k},k_z)= \sum_{l} \int \frac{dp_z}{ 2\pi} \int_{BZ}\frac{d^2p}{
 (2\pi)^2} {\hat b}_l^{\dagger}({\bs p},p_z) \frac{k_z}{m} {\hat b}_{l}({\bs p}+a\hat k,p_z+k_z)   \label{form-factor_{2t}}.
\end{align}
Because the form factor does not appear in the matrix element of the charge operator, the commutation relation between the charge density and electron operators 
becomes straightforward.  

The electron's propagator is defined as a matrix in the magnetic Brillouin zone and the Landau level index,
\begin{eqnarray}
\langle 0|{{\hat b}_m(\bs p,p_z) {\hat b}_n^{\dagger} (\bs p',p_z') }|0\rangle=(2 \pi)^4 \delta(\bs p-\bs p') \delta(p_z-p_z') {\hat S}_{mn}(\bs p,p_z).
\end{eqnarray} 
The vertex part is written as 
\begin{align}
\langle j^{\mu}(q) {\hat b}_m(p) {\hat b}_{n}(p') \rangle= (2\pi)^4 \delta(p+Q-p') {\hat S}_{mm'}(p,p+Q) {\hat \Gamma}^{\mu}_{m',n'} (p,p+Q) {\hat S}_{n'n} (p+Q),  
\end{align}
where $Q^{\mu}=(q_0,a \hat q_x,a \hat q_y,q_z)=t^{\mu}_{\nu} q^{\nu}$.
The vertex part  satisfied  the magnetic Ward-Takahashi identity 
\begin{eqnarray}
{\hat \Gamma}_{\mu}(p,p)=t_{\mu}^{\nu} \frac{\partial {{\hat S}(p)}^{-1}}{\partial p^{\nu}}, \label{W-T identity} 
\end{eqnarray}
and 
\begin{align}
\frac{\partial }{\partial Q_{\nu}} {\hat \Gamma}_{\mu} (p,p+Q)|_{Q=0}=
\begin{cases}
0;\ \nu=0,3;\ \mu=0,3 \\
\delta_{ij};\ \nu=i,\ \mu=j,\  i,j=1,2
\end{cases} \label{vertex_d}.
\end{align}
 \subsubsection{Neutrino-photon interaction}

Integrating over the electron fields in Eq. $(\ref{action})$, we obtain the effective Lagrangian for the neutrino and photon by applying derivative expansion on the fields $\tilde A_{\mu}(x) $ 
as   
\begin{eqnarray}
S_{eff}=\frac{1}{ 2} \int d^4x d^4y \tilde A_{\mu}(x) \pi^{\mu\nu}(x-y)
\tilde A_{\nu}(y),   \label{effective-action}
\end{eqnarray}
where  a Fourier transform of the current correlation 
function $\pi^{\mu\nu}(x-y)$   is expressed
\begin{align}
\pi_{\mu\nu}({\vec k})= \int \frac{d^4 p}{ (2\pi)^4}  \text{Tr}\left[{\hat \Gamma}_{\mu}(p,p+k){\hat S}(p+k){\hat \Gamma}_{\nu}(p+k,p)
{\hat  S}(p) \right],
\end{align}
where the sum in Tr is made in the Landau level index. 
Expansion coefficients of  $ \pi_{\mu\nu}({\vec k})$ in a power series on ${\vec k}$, give the strengths of the effective interaction. 
Among the coefficients, the linear term is most important in long-distance region, and its coefficient, i.e., the slope at the origin, is finite now for the space $ (\mu,\nu, \rho=0,x,y) $ 
due to the inversion symmetry breaking. 
This is expressed as a topological number of the propagator, using the Ward-Takahashi identity Eq. $( \ref{W-T identity})$,  
\begin{align}
&\frac{\partial}{ \partial k_{\rho}}\pi_{\mu\nu}(k)|_{k=0}=
 \epsilon_{\mu\nu\rho} \int \frac{d
 p_z}{  2\pi} \frac{1}{ 2\pi}N_{w} (p_z), \\
& N_w(p_z)=\frac{ \epsilon^{\mu\nu\rho}}{3!} 
\int \frac{d^3 p}{ (2\pi)^3} \text{Tr}\left[
{\hat S}_z(p) \left(\frac{\partial {\hat S_z(p)}^{-1}}{ \partial
 p_{\mu} }\right){\hat  S}_z(p) \left(\frac{\partial {\hat S_z(p)}^{-1}}{ \partial
  p_{\nu}} \right){\hat  S}_z(p) \left(\frac{\partial {\hat S_z(p)}^{-1}}{ \partial
  p_{\rho}}\right) \right] \label{topological_n}.
\end{align}
 where 
\begin{eqnarray}
\tilde S_z(p)={\hat  S}(p,p_z).
\end{eqnarray}
$N_w$ is the winding number of the propagator ${\hat S}(p,p_z) $ in $ (\mu,\nu, \rho=0,x,y) $ space-time. Its value, which depends on the magnetic field, is evaluated easily by substituting  
\begin{eqnarray}
\tilde S(\bs p,p_z)=U^{\dagger}(\bs p) \tilde S(\bs p,p_z) U(\bs p)
 \end{eqnarray} 
into Eq.$(\ref{topological_n})$.
$N_w(p_z)$ counts the number of Landau bands below the Fermi energy: 
\begin{eqnarray}
& &N_w(p_z)=\frac{\rho(p_z)}{\rho_0},\ \rho_0=\frac{eB}{ 2\pi \hbar},
\end{eqnarray} 
where  $\rho_0$ is the density of the Landau levels. 
This form is stable under perturbative corrections, and equivalent to the quantum Hall effect \cite{ishikawa, Coleman-Hill}. 


 \section{Anomalous neutrino-photon interaction}
\subsection{Chern-Simons form}

The effective action  Eq.$(\ref{effective-action})$ is decomposed to the lowest and higher dimensional terms:
\begin{align}
S_{eff} &=S_{CS}+S_{high}, \\
 S_{CS} &=\frac{\sigma_{xy}^{(4)}}{ 2}\int dz \int d^3 x 
 \epsilon^{\mu\nu\rho} {\tilde A}_{\mu}
 \partial_{\nu}{\tilde A}_{\rho}, \label{chern-simon} \\
\sigma_{xy}^{(4)} &=\frac{1}{ 2\pi} \nu^{(4)},\ \nu^{(4)}=\int \frac{dp_z}{ 2\pi} \nu(p_z)
= \frac{2\pi n_e}{  eB },
\end{align}
where $S_{high}$ is the higher dimensional term, $n_e$ is the electron density in the natural unit of $c=\hbar=1$.
The Lagrangian in $S_{CS}$ is proportional to the momentum and provides the dominant contribution in long-distance region. This system is invariant under a space time transformation that does not change the magnetic field. Accordingly, this is gauge and Lorentz invariant in the $(0,1,2)$ space, but not invariant under three-dimensional rotations, four-dimensional Lorentz transformations, and inversions. 


The effective Lagrangian in $S_{CS}$, the Chern-Simons (CS) term \cite{imos,iaim,ishikawa}, is
\begin{eqnarray}
\mathcal L_{int}= \frac{1}{2}\frac{\nu^{(4)}}{2\pi}\epsilon^{\alpha \beta \gamma} \tilde
 A_{\alpha} \partial_{\beta} \tilde A_{\gamma}  +O({\tilde
 F_{\alpha\beta}}^2);\ \alpha,\beta,\gamma=(0,1,2)
\label{effective-lagrangian},
\end{eqnarray}   
where the magnetic field is along the 3rd axis.  
 The vector potential includes the electromagnetic potential and neutrino current 
\begin{eqnarray}
& &\tilde A_{\alpha}=e A_{\alpha}(x)+\frac{G_F}{ \sqrt 2} \bar {\nu_e}(x)\gamma_{\alpha}(1-\gamma_5) {\nu_e}(x),
 \end{eqnarray}
and $ \tilde F_{\alpha\beta} =\partial_{\alpha}\tilde
A_{\beta}-\partial_{\beta} \tilde A_{\alpha}$. Due to this interaction term, the Maxwell equation is modified, and the electron neutrino interacts with the photon.  
The electromagnetic system in the $2+1$ dimensional space-time is that of a topologically massive gauge theory \cite{JDT}. 
The electron neutrino is the superposition of the mass eigenstates, ${\nu_e}(x)=\sum_i
U_{ei}{\nu_i}(x)$, with flavor eigenstates ${\nu_i}(x);\ i=1-3$ and  a mixing matrix $U$.  
The neutrino current is 
\begin{eqnarray}
& &J_{\alpha}(x)= g_{ij} \bar {\nu_i}(x) \gamma_{\alpha}(1-\gamma_5) {\nu_j}, \\
& &g_{ij}=U_{ei}^{*}U_{ej}. \nonumber
\end{eqnarray}

 The action describes the neutrino-photon interaction
\begin{eqnarray}
& &S_{\nu\gamma}=G_{\nu,\gamma}\int dz \int d^3 x 
 \epsilon^{\mu\nu\rho} J_{\mu}(x) 
 \partial_{\nu} A_{\rho}, \label{chern-simon2} \\
& & G_{\nu,\gamma}=\frac{e G_F}{\sqrt{2}} \frac{\nu^{(4)}}{ 2\pi}. \nonumber
\end{eqnarray} 
The coefficient $G_{\nu,\gamma}$, similar to the normal Hall and quantum Hall  effects in two-dimensional semi-conductors
\cite{klitzing,ishikawa}, is a
topologically invariant and satisfies   
a low energy theorem and remains the same in systems of
 disorders at finite-temperatures. The filling-factor $\nu^{(4)}$  is written in a manifestly  
Lorentz invariant expression 
\begin{eqnarray}
\nu^{(4)}=\frac{j^0}{B} \rightarrow \frac{j_{\mu} n^{\mu}}{\sqrt{F_{\mu \nu}F^{\mu \nu}}},
\end{eqnarray}
where $n^{\mu}=(1,0,0,0)$ in the rest system. 
\subsection{Physical effects of $ S_{{CS}}$  }
\subsubsection{Electromagnetic waves}

The action $ S_{CS} $ includes a  gauge invariant combination of  the vector potential.  
This term modifies  Maxwell equation:
\begin{eqnarray}
& &\frac{\partial S_{CS}}{\partial A_{\mu}(x)} =\sigma_{xy}^{(4)} \epsilon_{\mu\nu\rho} \partial_{\nu} A_{\rho}.
\end{eqnarray}
This shows that in a system of an electric field in the $x$-direction, an electric current flows in the $y$-direction, demonstrating the normal Hall effect.

Another effect of this term is a parity violating shift of the polarization direction of the photon, the Faraday rotation. This phenomenon was studied for  a quantum Hall system in \cite{ishikawa2,shimano}. 
\subsubsection{ Neutrino decays}
The action $S_{CS} $ is linear in momentum and is the lowest dimensional term.    
Other terms of higher power in the momentum are suppressed in the small momentum and  long-distance regions.   
The decays of the heavier neutrino to a lighter neutrino and a photon reveal the symmetric  properties of a system of the magnetic field, that is, the Lorentz invariance under the 2+1 dimensions, which violates  four dimensional Lorentz invariance.  In the process 
\begin{eqnarray}
\nu_i ({p}_{\nu_i}) \rightarrow \nu_f({ p}_{\nu_f})+\gamma( p_{\gamma}), \label{neutrino_decays}
\end{eqnarray}
the transition probability becomes a function of non-invariant quantities, 
\begin{align}
 p_{\nu_1}^z p_{\nu_2}^z, ~p_{\nu_i}^z p_{\gamma}^z;\ i=1,2. \label{m_variable}
\end{align}  

For particle decays that preserve energy and momentum and have four-dimensional Lorentz invariance, the bi-linear products of the momenta are reduced to  products of  masses.  
In Eq.$(\ref{neutrino_decays})$,
the three Lorentz invariant quantities, $p_{\nu_i}\! \cdot\! p_{\nu_f},\ p_{\nu_i}\!\cdot\! p_{\gamma},\ p_{\nu_f}\! \cdot\! p_{\gamma}$,
are not independent.  From energy-momentum conservation,  
\begin{eqnarray}
p_{\nu_i}=p_{\nu_f}+p_{\gamma}, 
\end{eqnarray}
and from mass shell conditions, $p_{\nu_i}^2=m_{\nu_i}^2,\ p_{\gamma}^2=m_{\gamma}^2$; these combinations are expressed as  
\begin{eqnarray}
& & p_{\nu_i} \cdot p_{\gamma}=\frac{1}{2}(m_{\nu_i}^2-m_{\nu_f}^2+m_{\gamma}^2), \label{mass_con}\\
& & p_{\nu_f} \cdot p_{\gamma}=\frac{1}{2}(m_{\nu_i}^2-m_{\gamma}^2-m_{\nu_f}^2),\\
& &p_{\nu_i} \cdot p_{\nu_f}=\frac{1}{2}(-m_{\gamma}^2+m_{\nu_i}^2+m_{\nu_f}^2), 
\end{eqnarray}
where $m_{\nu_i}$ and $m_{\gamma}$ are  the neutrino masses and effective mass of a photon in matter.
The Lorentz invariant bi-linear forms of the momenta are of order $m_{\nu}^2$ and $m_{\gamma}^2$ regardless of their energies. Even for the energy $E_{\nu_i}=1$ MeV, the square of the neutrino masses  are $\text{meV}^2$, which is  $10^{-18}$ times smaller than the square of the neutrino energy, $E_{\nu_i}^2$.     
Accordingly,  four-dimensional Lorentz invariant terms are negligible.    
As neutrinos have small masses, these terms are suppressed  even in high energy-region.

In contrast, the variables of Eq.$(\ref{m_variable})$ are not proportional to neutrino masses and the transition probability of   a massive neutrino to a neutrino and photon  that is derived from $S_{\nu \gamma}$ are not constrained by masses. 
\subsection{ Comparison of the electroweak Hall coupling with    the magnetic moment}
 
The effective action, Eq.$(\ref{chern-simon2})$, shows the neutrino's interaction with a photon  in a uniform magnetic field and electron density. This
is due to long-distance fluctuations and is not invariant under the four-dimensional Lorentz transformation. Its magnitude is independent of the neutrino masses.
$ S_{CS} $ is relevant in dilute plasma in a uniform magnetic field such as the solar corona and earth ionosphere. Using the magnetic field $B$ and electron density $n_e$ in the solar corona, 
\begin{eqnarray}
B^0=10^{-5}\ \text{T},  {n_e}^0=10^{15} \text{m}^{-3}, 
\end{eqnarray} 
with the strength
\begin{eqnarray}
G_{\nu,\gamma}=\frac{G_F }{\sqrt 2 }\frac{n_e}{n_e^0} \frac{B_0}{B}=6.58 \times 10^{-30} \frac{n_e}{n_e^0} \frac{B_0}{B}\ \text{MeV T}^{-1}.
\end{eqnarray}
This is much larger than the magnetic moment $\mu_{\nu}$ in a vacuum, as indicated by Eq.$(\ref{transition moment})$. $\mu_{\nu}$ in Eq.$(\ref{transition moment})$ is due to short-distance fluctuations and manifestly Lorentz invariant. The strength is proportional to the neutrino masses and extremely small.

 In a vacuum, stars, and other systems, the neutrino loses energy owing to its magnetic moment $\mu_{\nu}$. The current experimental upper bound 
\begin{eqnarray}
\mu_{\nu} \leq 0.022 \times 10^{-10} \mu_B,
\end{eqnarray} 
was derived from observations of the tip of the red giant in the globular cluster $\omega$-Centauri \cite{red-giant}.

 $G_{\nu, \gamma}$ appears in the system of the conduction electrons
 and uniform magnetic field. The coupling 
 strength is proportional to $\frac{n_e}{B}$ and not extremely
 weak even in a weak magnetic field if the electron coherence and uniformity of the magnetic field are kept in a large area.

 \subsection{ Wave packet decay : comparison with the plane wave decay }  
  
The wave of the finite coherence length is not a stationary state.  
A transition of non-stationary states is neither described by the stationary states nor by plane waves. These are described  by wave packets, characterized by a wavelength $l_w$ and  coherence length $l_c$. $l_w$ is the de Broglie wavelength $\frac{h}{p}=\frac{h c}{E}$ for a plane  wave of momentum $p$ and the energy $E=pc$. The overlapping area of the initial and final states is also stationary for plane waves; however, for moving wave packets, the overlapping  area is finite and moves. Integration over time reflects movement and shifts of the angular velocity from the stationary state. The value becomes  $\omega=\Delta E-{\vec V}^0 \cdot \Delta {\vec P}$, where $\Delta E$ is the energy difference, $\vec V_0$ is the velocity of the interaction area, and $\Delta \vec P$ is the difference of the momentum. For a particle of small mass, $m_l$,  $E_{l}-{\vec V_l}\cdot {\vec p}_{l}=\frac{m_l^2}{E_l}$ and is much smaller than $E_l$. Accordingly, the low-energy electron states in the intermediate states can contribute to the transition of much higher energy.  

\subsubsection{de Broglie wave length is the typical length }
 
We write the typical length of the physical system $l_{ph}$.  
For the scattering of the stationary state, the typical length is determined by the phase change over the space coordinate, and de Broglie wave length has a typical length of $l_{ph}=l_w$. 
The short-range correlation determines the probability and the scattering cross section  is governed by the length $l_w=\frac{h}{p}$. 
This has the atomic length for $p \geq 1$ keV, whereas the magnetic length is $R_{B}=\frac{mv}{eB}$, which is the length that  traps the electrons to the magnetic flux,  for $B=10^{-5}$ T and $1$ keV { $R_B\approx  1 $} m, which is much longer than the atomic distance. This does not affect the transition amplitude and probability in the short-distance region.  
The rates in this situation are the same as those in the vacuum, and are not modified by the     topological interaction Eq.$(\ref{chern-simon})$. As far as the phenomena caused by the rate are concerned, the topological interaction can be ignored.

\subsubsection{The typical length is larger than de Broglie wave length }
Real waves are not planes but  wave packets. Wave packets are normalized and have finite extensions.  
A transition probability expresses their transitions in a finite time interval $T$,  $P(T)$. 
In series of previous papers, $\cite{ishikawa-tobita-PTEP,ishikawa-tobita-ann,ishikawa-tajima-tobita} $, $\cite{ishikawa-oda,ishikawa-nishiwaki-oda1, ishikawa-nishiwaki-oda2}$, $P(T)$ was found to be expressed as 
 $P(T)=\Gamma T +P^{(d)}$, where $\Gamma$ is characterized by the de Broglie wave length, and $P^{(d)}$ is characterized by the coherent length $l_c$ of the wave function.
 Each wave packet propagates with a group velocity $v_g=\frac{\partial E(p)}{\partial p}$ and its phase varies according to the phase velocity  $v_{phase}=\frac{E}{p}$.   The change of the phase over the space and time coordinates determines the coherence length of the wave packet. This length is expressed by the difference of two velocities as $l_c=c \frac{h}{ p(v_g-v_{phase})}=ch \frac{E}{m^2c^4}$ for a particle of the small mass $m$ and the energy $E$. For light particles such as neutrinos, the difference between the two velocities is extremely small. This length is much longer than the de Broglie wave length. The previous assumption is invalid for the probability $P^{(d)}$.   

 Effects specific to these states appear in the physical length satisfying $l_{ph} \gg  R_{B}$ and are described by the action $ S_{CS} $; however, these  disappear in the short distance of $l_{ph}  \ll  R_{B}$.  
 $ S_{CS} $ is the action of the lowest dimension and reveals the physical length $l_{ph} \gg R_{B}$.

 The long-range property of $P^{(d)}$ is similar to that of the  EPR correlation.  
The coherence length $ l_{c} = \frac{c\hbar E }{m^2c^4}$ for the neutrino, the mass is of meV and the energy is of order MeV, can become of order $10^5$ km for $p\geq 1$ MeV.  In previous papers \cite{ishikawa-tobita-PTEP,ishikawa-tobita-ann,ishikawa-tajima-tobita}, it was shown that an angular frequency of the interference term of two nearby masses is determined by the difference $\frac{\delta m^2c^2}{p}$.  
For the mass range of meV and momentum of MeV, this length satisfied $l_{ph} \gg R_{B} $ and the system is governed by the topological interaction Eq. $(\ref{chern-simon})$. 

The interaction Eq.$(\ref{chern-simon2})$  applies to relativistic electrons and in a finite temperature system. Neutrinos,  radiation, and electrons in a magnetic 
field interact in the solar corona. In the quiet corona, $B=10^{-4} \text{ T}$ and $n_e= 10^{11}-10^{15}\text{ m}^{-3}$, and the number of the  cyclotron oscillation within a mean free path is $\omega_B \times \frac{l_{\text{mfp}}^e}{ v} \gg 1 $, where $\omega_B=\frac{eB}{  m_e}$, $l_\text{mfp}^e$ and $v$ are the mean free path and velocity of electrons. Thus, the electron motion is determined by the electrons in
 the external magnetic field and the long-distance fluctuation is
 described by the eigenmodes in an external  magnetic field.

\section{Summary and prospects}
The electroweak Hall Lagrangian of the neutrinos and photon fields in the dilute magnetized plasma is obtained using the electron propagator in the magnetic field.  This treatment fully considers non-perturbative quantum effects  due to  electron fluctuations. 

The effective action Eq. $( \ref{chern-simon2})$ and those of the bi-linear form of the fields  represent the system's dynamics, that is, the propagation and interaction of the fields. The bi-linear form of the fields are practically the same as those in the vacuum. The electroweak Hall  Lagrangian shows a new interaction between a neutrino and photon in magnetized plasma.
This  causes an electroweak Hall effect between the photon and a pair of neutrinos with the  same origin as the electronic  Hall effect in the semiconductor. 
The coupling strength is proportional to the topological invariance of the electron propagator in the Landau levels $\frac{n_e}{B} \times eG_F$ and is inversely proportional to the magnetic field. This interaction applies to the colored regions  in Figs. 1--3.

  The magnetic field is  a Lorentz tensor, which is transformed  to another system of magnetic and electric fields by a Lorentz transformation. From a perspective of a moving observer, the  system has an electric and magnetic field. The effective action Eq. $( \ref{chern-simon2})$ represents the interaction of the fields in the presence of a uniform magnetic field in the $z$-direction. This depends on the direction and is not manifestly invariant under Lorentz transformation in  $3+1$ dimensions.

The radiative transitions of the neutrino are derived from the electroweak Hall Lagrangian. Its implications at large distance regions will be presented in a future paper \cite{corona}.

\subsection*{Acknowledgments}
 This work was partially supported by a 
Grant-in-Aid for Scientific Research ( Grant No. JP21H01107).

\appendix 
\section{ Landau levels of Dirac equation}
In the following the unit $c=1$ is used.
\begin{align}
( i \hbar \frac{\partial}{\partial t} -m)\phi &= \vec  \sigma \cdot( \vec   p+e \vec  A)\xi, \\
 (i \hbar \frac{\partial}{\partial t} +m) \xi &= \vec  \sigma \cdot( \vec  p+e \vec  A)\phi. 
\end{align}
The stationary solutions $\phi=e^{i \frac{E}{\hbar} t} \phi,\ \xi=e^{i \frac{E}{\hbar} t} \xi $
\begin{align}
(E  -m)\phi &= \vec  \sigma \cdot( \vec  p+e \vec  A)\xi,   \\
(E +m) \xi &= \vec  \sigma \cdot( \vec  p+e \vec  A)\phi, 
\end{align}
where
\begin{align}
(E^2  -m^2)\phi &=\left[  {\vec  p}^2 +e^2 {\vec  A}^2 +2 e \vec  A \cdot \vec  p   +e \vec  \sigma \cdot \vec  B \right] \phi, \\
E&=\pm\sqrt{m^2+p_z^2+(2l+1+\sigma)\hbar eB  }.
\end{align}
In the non-relativistic region, the energy is expressed as
\begin{align}
E=\pm \left[ m+ \frac{p_z^2+(2l+1+\sigma) eB }{2m}+\text{higher orders}\right].
\end{align}
The spin is factorized from an orbital motion.

\end{document}